\documentstyle[10pt]{article}
\begin{document}
\begin{center}
{\large\bf Probing Multiparton Correlations at CEBAF}
\medskip

Jianwei Qiu\footnote{On-leave from Department of 
Physics and Astronomy, Iowa State University,
Ames, Iowa 50011, USA.}\\
{\it Physics Department, Brookhaven National Laboratory}\\
{\it Upton, New York 11973-5000, USA}
\end{center}

\begin{abstract}
In this talk, I explore the possibilities of probing
the multiparton correlation functions at CEBAF at its 
current energy and the energies with its future upgrades.
\end{abstract}
\medskip

\centerline{\bf I. INTRODUCTION}
\smallskip

The perturbation theory of Quantum Chromodynamics (QCD) 
has been very successful in interpreting the data from high 
energy experiments.  For observables involving hadrons, 
QCD factorization theorem enables us to 
systematically separates the short-distance partonic dynamics
from the long-distance non-perturbative hadronic informations.
In particular, when the energy exchange in a collision 
is high enough, the QCD factorization theorem factorizes all
non-perturbative information into a set of universal parton 
distributions.  Such universal parton distributions have been 
successfully extracted from the existing data.

However, the nonperturbative dynamics of the strong interactions 
is much richer than what we have learned through 
the parton distributions.  The parton distributions are 
interpreted as the 
probability densities to find the partons within a hadron.  
Knowing only such probability densities is not enough
to understand the full partonic dynamics within a hadron.  
But, very little has been learned  
about the multiparton correlations.  This is because 
the physics associated with the multiparton 
correlations is directly related to multiple scattering, and 
most physical observables measured in existing high 
energy experiments are dominated by the contributions from the
single hard scattering.  And, in general, the multiple scattering 
is power suppressed in comparison with the single scattering.
Therefore, it is often the case that even an excellent data 
cannot provide a good enough information for extracting the 
multiparton correlations, after removing the large contributions
from the single scattering.   

In this talk, I use examples to show that (1) by
taking the advantage of kinematics, we can enhance the contribution 
from the multiparton scattering; and (2) there are new types of 
physical observables which eliminate the contribution from the 
single scattering.  The rest of this talk is organized as follows.
A general discussion on relationships between the multiple scattering,
the multiparton correlation functions and the high twist operators is 
given in next section.  In Sec.~III, with some simple assumptions,
we show analytically that when $x_B$ is large,
the complete twist-4 contributions to the DIS structure functions 
can be simplified and is proportional to the derivative of the 
twist-2 parton distributions. 
In Sec.~IV, we demonstrate how to extract multiparton correlation
functions inside a large nucleus.  Finally, in Sec.~V, 
we provide our summary and discussion on why CEBAF is a good place
to measure the multiparton correlation functions.

\medskip
\centerline{\bf II. MULTIPARTON CORRELATIONS AND HIGH TWIST}
\smallskip

Multiparton correlation functions are defined as the Fourier 
transform of the matrix elements of the multiparton fields,  
\begin{eqnarray}
T_m(k_1,k_2,\dots,k_{n-1}) 
&\propto &\int\,
\prod_{m=1}^{n-1} \frac{d^4y_m}{(2\pi)^4}\
e^{ik_m\cdot y_m}
\nonumber \\
&\times &
\langle N|\phi(0)\phi(y_1)\dots\phi(y_{n-1})|N\rangle \, ,
\label{Tn}
\end{eqnarray}
where $|N\rangle$ is the hadron state, the $k_i$ with $i=1,..,n-1$ 
are independent parton momenta, and the $\phi$'s are 
the parton field operators.  Because of the momentum 
dependence of the correlation functions, the multiparton 
field operators in the position space, $\phi(0)\phi(y_1)
\dots \phi(y_{n-1})$ are not local.  In a gauge theory,
like QCD, a path integral of the gauge field is often necessary 
to be sandwiched between the $\phi$'s in Eq.~(\ref{Tn}), in order
to make the correlation functions gauge invariant \cite{QS_DIS}.  
The simplest correlation functions are the matrix elements of 
operators with two parton fields.  For example, the quark 
distribution of a proton of momentum $p$, moving in the
``+'' direction,
\begin{equation}
q(x)=\int\frac{dy^{-}}{2\pi}e^{ixp^{+}y^{-}}
\langle p|\bar{\psi}_q(0)\left(\frac{\gamma^{+}}{2}\right)
               \psi_q(y^{-})|p\rangle \, ,
\label{qx}
\end{equation}
which is a correlation of two quark fields.  

Beyond two-parton correlation functions, the choice of the 
multiparton field operators is not unique because of the 
spin structure of the quark fields, the equation of motion, 
and the gauge invariance.  For example, the structure
functions measured in deeply inelastic scattering (DIS)
can have following factorized form,
\begin{equation}
F(x_B,Q^2)=\sum_m C^{(m)}(x_B,x_1,x_2,..,Q^2/\mu^2)
\otimes T^{(m)}(x_1,x_2,..,\mu^2)\left(\frac{1}{Q^2}\right)^m,
\label{fxb}
\end{equation}
where $x_B$ is the Bjorken variable, and $Q^2$ is the large
virtual momentum exchange in DIS.  In Eq.~(\ref{fxb}), the 
coefficient functions, $C^{(m)}$'s, are calculable within the QCD
perturbation theory, and the $T^{(m)}$'s are nonperturbative and 
defined as the Fourier transform of the matrix elements of the
multiparton field operators.  The term with $m=0$ is often called 
the leading twist or leading power contribution, while
the terms with $m\neq 0$ are called the high twist contributions 
or power corrections.  It was pointed out in Ref.~\cite{Qiu} that
at a given order of power corrections, $m$, the $T^{(m)}$ can be
expressed in terms of the multiparton correlation functions with 
the same number of physical parton fields (i.e., fields with 
physical polarizations).  For example, the $T^{(1)}$ for the 
$m=1$ power corrections in DIS can be expressed in terms of 
{\it four}-parton correlation functions only (note, the power
corrections due to the hadron mass are not included in 
discussion here).  

Although the {\it four}-parton correlation functions correspond to 
the {\it leading} power corrections to the structure functions in 
DIS or other physical observables, we have not had much success 
in extracting the information on these correlation functions.  
This is because of following obvious difficulties:
\begin{itemize}
\item The twist-4 (or higher twist) contributions are power suppressed
in comparison with the corresponding leading twist contributions.
\item There are many more four-parton correlation functions than 
the two-parton correlation functions (or parton distributions) due
to the rich spin decompositions and flavor combinations of different 
parton field operators.  
\item Dependence on extra parton momenta requires much more 
information to extract multiparton correlation functions
than the parton distributions.
\end{itemize}
In order to extract the multiparton correlation functions, we have to 
identify the physical observables such that above difficulties can
be minimized.  For example, we can measure a physical observable in 
a particular part of the phase space so that the power corrections are 
relatively enhanced; we can identify 
an observable which gets no leading twist contribution; 
we can take an advantage of kinematics
and identify observables with a limited leading subprocesses 
so that only a small number of multiparton correlation functions 
are relevant \cite{LQS,Guo}. 
In the rest of this talk, we will give two examples to show how
to extract the four-parton correlation functions, and provide
the physical interpretation of these four-parton correlation 
functions.

\medskip
\centerline{\bf III. TWIST-4 CONTRIBUTION TO THE STRUCTURE}
\centerline{\bf IN DEEPLY INELASTIC SCATTERING}

In this section, we demonstrate how to extract the leading power 
corrections (i.e., the twist-4 contributions) to the structure 
functions in DIS at the CEBAF energies.  In particular, we provide 
the physical interpretation of the leading power corrections.
Note, the inclusion of target mass corrections are straightforward, 
and will not be discussed here.

Although the twist-4 contributions to the structure functions in 
DIS are power suppressed by a factor of $(1/Q^2)$, as shown in 
Eq.~(\ref{fxb}), the corresponding coefficient functions
$C^{(1)}(x_B,x_1,x_2,\dots)$ are larger than the leading 
$C^{(0)}(x_B)$'s for large $x_B$.  The structure functions 
at large $x_B$ are useful for extracting
the information on the size of power corrections \cite{Bodek}.  
However, after removing the leading twist contributions, the 
existing data are still not good enough for extracting detailed 
information on different four-parton correlation functions
$T^{(1)}$'s in Eq.~(\ref{fxb}).  It is a common practice to 
use the following parameterization for the leading power 
corrections \cite{Bodek},
\begin{equation}
F_2(x_B,Q^2) \approx \left(1+\frac{h(x)}{Q^2} \right)
F_2(x_B,Q^2)_{{\rm LT}}\, ,
\label{f2}
\end{equation}
where $F_2(x_B,Q^2)_{{\rm LT}}$ includes the leading twist
(or leading power) contribution and the target mass corrections to 
the full structure function $F_2(x_B,Q^2)$.  The $h(x)$ in
Eq.~(\ref{f2}) is a fitting function. Yang et al. \cite{Bodek} 
used following parameterization, 
\begin{equation}
h(x_B) = a \left(\frac{x_B^b}{1-x_B} - c\right) \, ,
\label{hx}
\end{equation}
to fit the existing DIS data and extract the values of the parameters
$a,b$ and $c$ for different targets.  
On the other hand, a complete expression of the twist-4 contributions 
to the DIS structure functions at the leading order of $\alpha_s$ 
were derived more than fifteen years ago \cite{Qiu,Ellis}.  It is 
important to find the linkage between the phenomenological 
fits and the theoretical formulas, and
reveal the physical meanings of the these fitting parameters.

In terms of one photon exchange, the invariant mass square of the 
hadron-photon system is given by
\begin{equation}
W^2 = (p+q)^2 
\approx \frac{Q^2}{x_B} \left(1-x_B\right) \, .
\label{W2}
\end{equation} 
For inclusive DIS, it is obvious that the power expansion of the 
structure functions in Eq.~(\ref{fxb}) should have terms
proportional to $(1/W^2)^m$.  Because $W^2$ shown in Eq.~(\ref{W2}) 
can be much smaller than $Q^2$ in large $x_B$ region, the power 
corrections are relatively more important in this region.
Without worrying about the target mass corrections, 
the coefficient functions should have following asymptotic 
behavior,
\begin{equation}
C^{(m)}(x_B,x_1,x_2,\dots,Q^2/\mu^2) 
\Longrightarrow \left(\frac{1}{1-x_B}\right)^m\, 
\end{equation}
as $x_B\rightarrow 1$.  From above simple argument based on the DIS 
kinematics, it seems natural to parameterize the fitting function 
$h(x_B)$ in Eq.~(\ref{hx}).

In addition to the asymptotic behavior of the coefficient 
functions, it is also important to understand why the the twist-4 
part of the structure function, $F_2(x_B,Q^2)_{\rm TW4}$, which are 
proportional to the four-parton correlation functions, can be 
approximated to be proportional to the twist-2 part of the 
structure functions, as shown in Eq.~(\ref{f2}).  The 
twist-4 contributions to the structure functions at the 
leading order of $\alpha_s$ are given by \cite{Qiu,Ellis}
\begin{eqnarray}
F_2(x_B,Q^2)_{{\rm TW4}} &=& \frac{x_B}{Q^2} \sum_q e_q^2\, \Bigl[
4\, T_{qD_1}^{(1)}(x_B) - x_B\int dx_1 dx_2 
\nonumber \\
&& \times 
\frac{\delta(x_2-x_B)-\delta(x_1-x_B)}{x_2-x_1}\,
T_{qD_2}^{(1)}(x_2,x_1)\Bigr] \, ;
\label{F2TW4}
\end{eqnarray}
and
\begin{equation}
F_L(x_B,Q^2)_{{\rm TW4}} = \frac{1}{Q^2} \sum_q e_q^2\, \Bigl[
4\, T_{qD_1}^{(1)}(x_B) \Bigr] \, ,
\label{FLTW4}
\end{equation}
plus the contributions from the four-quark correlation functions.
In Eqs.~(\ref{F2TW4}) and (\ref{FLTW4}), the quark-gluon correlation
functions are defined as
\begin{equation}
T_{qD_1}^{(1)}(x_B) = \frac{1}{4} \int \frac{dy^-}{2\pi}\,
e^{ix_Bp^+y^-} \langle p|\bar{\psi}_q(0)
\gamma_\alpha\gamma^+ \gamma_\beta
D^\alpha_T(0)D^\beta_T(y^-)\psi_q(y^-) |p\rangle \, ;
\label{T41}
\end{equation}
and
\begin{eqnarray}
T_{qD_2}^{(1)}(x_B) &=& \frac{1}{4} \int 
\frac{dy^-}{2\pi}\, \frac{p^+dy_1^-}{2\pi}\,
e^{ix_1p^+y^-}\, e^{i(x_2-x_1)p^+y_1^-} \nonumber \\
&& \times 
\langle p|\bar{\psi}_q(0)
\gamma_\alpha\gamma^+ \gamma_\beta
D^\alpha_T(y_1^-)D^\beta_T(y_1^-)\psi_q(y^-) |p\rangle\, ,
\label{T42}
\end{eqnarray}
where $D_T$ are transverse components of the covariant derivatives.

In Ref.~\cite{GQZ}, we show that when $x_B$ is large, a simplified 
expression, similar to the phenomenological parameterization, can be
derived from Eq.~(\ref{F2TW4}).
Key arguments and 
assumption of the derivation are summarized as follows:
\begin{itemize}
\item It is argued that the correlation function $T_{qD_1}^{(1)}(x_B)$
cannot be proportional to $F_2(x_B,Q^2)_{\rm LT}/(1-x_B)$ as 
$x_B\rightarrow 1$.  Therefore, the leading contribution to 
$F_2(x_B,Q^2)_{\rm TW4}$ is from the second term in Eq.~(\ref{F2TW4}).
\item Assuming that $D_T^2(y_1^-)\equiv -g_{\alpha\beta}\,
D_T^{\alpha}(y_1^-)D_T^{\beta}(y_1^-)$ is a very slow function of 
$y_1^-$, we can reduce 
the quark-gluon correlation function $T_{qD_2}^{(1)}$
to the normal twist-2 quark distribution defined
in Eq.~(\ref{qx}).
\item It is argued that the contributions from the four-quark 
correlation functions are smaller than that from the quark-gluon
correlation functions given in Eq.~(\ref{F2TW4}).
\end{itemize}
With the above arguments and assumption, we derive 
\begin{eqnarray}
F_1(x_B,Q^2)_{{\rm TW4}} &=& \frac{1}{2}\left[
\frac{F_2(x_B,Q^2)_{\rm TW4}}{x_B} - F_L(x_B,Q^2)_{\rm TW4}\right]
\nonumber \\
&\approx & \frac{1}{2}
\sum_q e_q^2\, \frac{\langle D_T^2 \rangle}{2Q^2}\,  
x_B \frac{d}{dx_B} q(x_B)\, ,
\label{F1}
\end{eqnarray} 
where the twist-2 quark distribution $q(x_B)$ is defined in 
Eq.~(\ref{qx}).  The $\langle D_T^2 \rangle$ is an averaged
value of the covariant derivative.  Eq.~(\ref{F1}) is a direct
result of above arguments and assumption.  As shown below,
Eq.~(\ref{F1}) can be reduced to an expression very similar 
the phenomenological parameterization given in Eqs.~(\ref{f2})
and (\ref{hx}).

Eq.~(\ref{F1}) in its present form is still different from 
the parameterization shown in Eq.~(\ref{hx}) because of the 
derivative of the quark distribution.  However, when $x_B$ 
is large, the quark distribution should have following 
behavior
\begin{equation}
q(x_B) \Longrightarrow (1-x_B)^{\alpha_q}
\quad \mbox{as $x_B\rightarrow 1$} \, , 
\end{equation}
with the real parameter $\alpha_q \sim 3.0$ for the valence 
quark distributions.  Therefore, we have
\begin{equation}
x_B \frac{d}{dx_B} q(x_B) =
\alpha_q\, \frac{x_B}{1-x_B}\, q(x_B)\,
+\, \mbox{terms without}\ \frac{1}{1-x_B}\, .
\label{dxqx}
\end{equation}
Substituting Eq.~(\ref{dxqx}) into Eq.~(\ref{F1}), we obtain
\begin{eqnarray}
F_2(x_B,Q^2)_{\rm TW4} &\approx & 
\sum_q e_q^2 \, x_B\, q(x_B) \,
\frac{\alpha_q \langle D_T^2 \rangle}{2Q^2}\, 
\frac{x_B}{1-x_B}
\nonumber \\
&& +\ \mbox{terms without}\ \frac{1}{1-x_B}\, ,
\label{F2TW4a}
\end{eqnarray}
It is now clear that our analytical result shown in Eq.~(\ref{F1}) or 
equivalently Eq.~(\ref{F2TW4a}) is very similar to 
the phenomenological parameterization shown in Eqs.~(\ref{f2}) 
and (\ref{hx}).  The difference is that our result was derived 
analytically from the complete leading order twist-4 contributions
with some simple arguments and assumptions, and thus,
every parameter have the physical interpretations.  In addition,
because of the dependence on the derivative of the quark 
distributions, instead of the leading twist structure function,
our result provides the natural difference between the proton 
target and the neutron target, which is observed in the NMC data.

In order to measure the power corrections to the structure functions,
we like to keep $x_B$ large and $Q^2$ small.  However, 
from Eq.~(\ref{W2}), we cannot keep $Q^2$ too small, because of 
the hadronic resonances when $W$ is less than 2~GeV.  When 
the invariant mass of the ``hadron-photon'' system is less 
than 2~GeV, the power expansion shown in Eq.~(\ref{fxb}) fails.
Therefore, to test the perturbative power expansion of the
structure functions and to measure the multiparton correlation
functions, it is important to keep $W\geq 2$~GeV.
With its luminosity and energy (in particular, with its future 
upgrades), CEBAF should be a good place to measure such  
power corrections, and to test the result shown in Eq.~(\ref{F1}).

\medskip
\centerline{\bf IV. MULTIPARTON CORRELATION IN A NUCLEUS}
\smallskip

Multiparton correlation functions inside a large nucleus
are extremely important and useful for understanding nuclear
dependence in relativistic heavy ion collisions.  Inside a 
large nucleus, multiple scattering, which is directly
associated with the multiparton correlation functions, can take
place within one nucleon or between different nucleons.  Since 
the leading single scattering at high energy is always localized
within one nucleon, it cannot generate any large dependence on 
nuclear size.  Similarly, multiple scattering within one nucleon
cannot provide much dependence on nuclear size either.  Therefore,
large dependence on nuclear size is an unique signal of multiple
scattering between nucleons.  Measurement of such anomalous 
dependence on the nuclear size for any physical observable
will provide direct information on the multiparton correlation 
functions in a nucleus.  However, in order to test QCD, and 
to extract the useful information on the multiparton correlation
functions, it is important to identify the physical observables
which depend only on a small number of correlation functions.
Otherwise, the data will not be able to separate contributions 
from different multiparton correlation functions.

It was pointed out recently in Ref.~\cite{Guo2} that at the leading 
order of $\alpha_s$, it needs only one type of multiparton 
correlation function for both transverse momentum broadening of 
Drell-Yan pair and the jet broadening in DIS.  It is a
quark-gluon correlation function, and is defined as 
\begin{eqnarray}
T_{qF}^{A}(x_B)&=& \int \frac{dy^-}{2\pi} 
e^{ix_B p^+ y^-} \frac{dy_1^- dy_2^-}{2\pi}
\theta(y^--y_1^-)\theta(-y_2^-)
\nonumber \\
&\times &
\langle P_A| F^+_{\ \alpha}(y_2^-) 
\bar{\psi}(0) \frac{\gamma^+}{2} \psi(y^-)
F^{+\alpha}(y_1^-)|P_A\rangle \, ,
\label{T4qA}
\end{eqnarray}
with $A$ the atomic weight.
In terms of this correlation function, the jet broadening in DIS
can be expressed as \cite{Guo2}
\begin{equation}
\Delta\langle p_T^2\rangle = \frac{4\pi^2\alpha_s}{3}\,
\frac{\sum_q e_q^2\,T_{qF}^{A}(x_B)}{\sum_q e_q^2\,q^{A}(x_B)}\, ,
\label{jetdis}
\end{equation}
where $p_T$ is the transverse jet momentum in the photon-hadron 
frame in DIS.  Measuring the jet momentum broadening, 
$\Delta\langle p_T^2\rangle$,
provides a direct measurement of the quark-gluon correlation
functions inside a nucleus.

However, at the CEBAF energies, it is not possible to measure the
jet cross section and corresponding jet momentum broadening in
nuclear targets.  But, it should be possible
to measure the transverse momentum broadening of the leading pions
(e.g., final-state pions with a relatively large momentum).  Define
the averaged transverse momentum of the pions of momentum $\ell$ as
\begin{equation}
\langle \ell_T^2\rangle^{eA} = \frac{\int d\ell_T^2 \, \ell_T^2 \, 
\frac{d\sigma_{eA\rightarrow \pi}}{dx_B  dQ^2 d\ell_T^2} }
{\frac{d\sigma_{eA}}{dx_B  dQ^2} }\, ;
\end{equation}
and define the nuclear broadening as
\begin{equation}
\Delta\langle \ell_T^2\rangle = 
\langle \ell_T^2\rangle^{eA} - \langle \ell_T^2\rangle^{eN}\, .
\end{equation}
We obtain \cite{GQ} that
\begin{equation}
\Delta\langle \ell_T^2\rangle = 
\frac{4\pi^2\alpha_s}{3}\,
\frac{\sum_q e_q^2\,\int_{z_{\rm min}}^1 
             dz\, z^2 D_{q\rightarrow\pi}(z) T_{qF}^{A}(x_B)}
     {\sum_q e_q^2\,q^{A}(x_B)}\, ,
\label{pilt}
\end{equation}
where $D_{q\rightarrow\pi}(z)$ are the quark-to-pion fragmentation
functions, and $z_{\rm min}$ depends on minimum pion momentum
and is frame dependent.  Since the quark-to-pion fragmentation 
functions are known, Eq.~(\ref{pilt}) provides the direct 
information on the quark-gluon correlation functions inside a 
nucleus.  By measuring the transverse momentum broadening for 
both $\pi^{\pm}$ and $\pi^0$, and keep a reasonable large
value of $z_{\rm min}$, we can extract the quark flavor 
dependence of the correlation functions.

\medskip
\centerline{\bf V. CONCLUSIONS AND OUTLOOK}
\smallskip

We present an analytical derivation of the leading
twist-4 contributions to the DIS structure functions when $x_B$ is 
large, and argue that CEBAF is a good place to test such 
contributions.  Our derivation provides a direct linkage between the 
complete twist-4 contributions and the phenomenological 
parameterization used for fitting the existing data; and physical 
meanings of the parameters in the phenomenological parameterization.  
Besides the similarities between the analytical result and the
parameterization, there are also some differences.  It is the 
difference that makes the measurements at CEBAF even more interesting.

We also demonstrated that measuring the transverse momentum 
broadening of the leading pion production at CEBAF can provide
direct information on multiparton correlation functions in a large 
nucleus.  Such information are extremely important for the multiple
scattering and anomalous nuclear dependence in relativistic heavy
ion collisions.


\end{document}